\documentclass[aps,prb,reprint,groupedaddress,footinbib,citeautoscript,showpacs]{revtex4-1}

\usepackage{graphicx}
\usepackage{amsmath}
\usepackage{amssymb}
\usepackage{bm}
\usepackage{array}
\usepackage{cleveref}
\newcolumntype {s}[1]{@{\hspace{#1}}} % space
\usepackage{color}
\usepackage{wrapfig}

\graphicspath{{.}{./FinalFigures/}{./EPS/}}

\newcommand{\ket}[1]{\left | \, #1 \right \rangle}

\newcommand{\av}[1]{\langle #1\rangle}

\newcommand* {\ee}{\mathrm{e}}
\newcommand* {\Ee}{\mathcal{E}}
\newcommand*{\vek}[1]{{\bm{\mathrm{#1}}}}

\newcommand*{\kk}{{\bm{\mathrm{k}}}}

\newcommand*{\pp}{{\bm{\mathrm{p}}}}

\newcommand{\pvec}[1]{\vec{#1}\mkern2mu\vphantom{#1}}
\newcount\colveccount
\newcommand*\colvec[1]{
        \global\colveccount#1
        \begin{pmatrix}
        \colvecnext
}
\def\colvecnext#1{
        #1
        \global\advance\colveccount-1
        \ifnum\colveccount>0
                \\
                \expandafter\colvecnext
        \else
                \end{pmatrix}
        \fi
}

\def\lsim{\raise0.3ex\hbox{$\;<$\kern-0.75em\raise-1.1ex\hbox{$\sim\;$}}}
\def\gsim{\raise0.3ex\hbox{$\;>$\kern-0.75em\raise-1.1ex\hbox{$\sim\;$}}}

\DeclareMathSymbol{\myRe}{\mathord}{symbols}{"3C}

\DeclareMathSymbol{\myIm}{\mathord}{symbols}{"3D}

\begin{document}

\title{Coulomb-exchange effects in nanowires with spin splitting due to a radial electric field}

\author{F. S. Gray}
%\email{grayfinn.@myvuw.ac.nz}
\affiliation{School of Chemical and Physical Sciences and MacDiarmid
Institute for Advanced Materials and Nanotechnology, Victoria
University of Wellington, PO Box 600, Wellington 6140, New Zealand}

\author{T. Kernreiter}
%\email{thomas.kernreiter@vuw.ac.nz}
\affiliation{School of Chemical and Physical Sciences and MacDiarmid
Institute for Advanced Materials and Nanotechnology, Victoria
University of Wellington, PO Box 600, Wellington 6140, New Zealand}

\author{M. Governale}
\email{michele.governale@vuw.ac.nz}
\affiliation{School of Chemical and Physical Sciences and MacDiarmid
Institute for Advanced Materials and Nanotechnology, Victoria
University of Wellington, PO Box 600, Wellington 6140, New Zealand}

\author{U. Z\"ulicke}
\email{uli.zuelicke@vuw.ac.nz}
\affiliation{School of Chemical and Physical Sciences and MacDiarmid
Institute for Advanced Materials and Nanotechnology, Victoria
University of Wellington, PO Box 600, Wellington 6140, New Zealand}

\date{\today}

\begin{abstract}

We present a theoretical study of Coulomb exchange interaction for electrons
confined in a cylindrical quantum wire and subject to a Rashba-type spin-orbit coupling
with radial electric field. The effect of spin splitting on the single-particle band
dispersions, the quasiparticle effective mass, and the system's total exchange energy per
particle are discussed. Exchange interaction generally suppresses the quasiparticle effective
mass in the lowest nanowire subband, and a finite spin splitting is found to significantly increase
the magnitude of the quasiparticle-mass suppression (by upto 15\% in the experimentally
relevant parameter regime). In contrast, spin-orbit coupling causes a modest (1\%-level)
reduction of the magnitude of the exchange energy per particle. Our results shed new light
on the interplay of spin-orbit coupling and Coulomb interaction in quantum-confined systems,
including those that are expected to host exotic quasiparticle excitations.

\end{abstract}


%\keywords{}

\maketitle

\section{Introduction}

The dimensionality of a many-particle system is a crucial determinator for how importantly
interaction effects can shape its physical properties. Generally, three-dimensional (3D)
bulk conductors are less drastically affected by the Coulomb interaction between charge
carriers than lower-dimensional, quantum-confined structures such as quasi-2D quantum 
wells and quasi-1D quantum (nano-)wires~\cite{Giuliani2005}. This is essentially due to
phase-space restrictions arising from free motion being only possible in fewer than three
spatial directions. Furthermore, the exact structure of transverse bound-state wave
functions shapes the density distribution of the confined charge carriers and, thus, turns
out to critically influence Coulombic effects in quantum wells~\cite{Ando1982} and
wires~\cite{Gold1990}. Here we explore how another aspect of quantum-confined states,
namely their intrinsic spinor structure, modifies the effect of the Coulomb interaction  
in nanostructured systems.

Most low-dimensional conductors are fabricated from semiconductor materials where the
coupling between the spin degree of charge carriers and their orbital motion is often quite
strong~\cite{Yu2010}. As a result, quantum confinement can significantly affect spin-related
properties~\cite{Winkler2003}. Such effects are particularly
pronounced for valence-band states (i.e., \emph{holes\/}) because of their peculiar
spin-$3/2$ character~\cite{Winkler2003,Winkler2005}. In contrast, conduction-band
electrons are spin-$1/2$ particles and generally subject to weaker spin-orbit couplings that
are due to the bulk inversion asymmetry in the material's crystallographic unit cell
(Dresselhaus~\cite{Dressel1955}  spin splitting) or the structural inversion asymmetry
present in a nanostructured systems (Rashba~\cite{Rashba1960,Bychkov1984} spin
splitting). The multitude, and often counter-intuitive nature, of spin-orbit effects in
nanostructures has become the focus of recent study, with developing an understanding of
the interplay with Coulomb interactions being a key question to be addressed. Bulk-hole
systems~\cite{Comb1972,Schlie2006,Schlie2011,Kyry2011}, quantum-well-confined
holes~\cite{Schmitt1994,Cheng2001,Chesi2007,Scholz2013,Kernreiter2013}, and 2D
electron systems subject to Rashba spin
splitting~\cite{Pletyukhov2006,Chesi2007,Chesi2011,Chesi2011a,Agarwal2011} have
been considered. The comparatively few studies of Coulomb-interaction effects in
spin-orbit-coupled quasi-1D
systems~\cite{Haeusler2001,Gritsev2005,Schulz2009,Maier2014} have almost exclusively
focused on effective Luttinger-liquid descriptions~\cite{Giamarchi2004}
and, in particular, did not investigate the effect of Rashba spin splitting on the total exchange
energy and exchange-induced quasiparticle-effective-mass renormalization in quantum  wires.

In this article, we fill precisely this gap and investigate both the exchange energy and effective-mass 
renormalization in quantum wires with a Rashba-type spin-orbit coupling.
Previous work on the exchange energy of quantum wells revealed that spin-orbit coupling has the
opposite effect on interactions in n-type and p-type systems: the exchange energy of a
quasi-2D conduction-band electron system is slightly enhanced~\cite{Chesi2007,Chesi2011a}
due to spin-orbit coupling, whereas the exchange energy for quasi-2D holes is suppressed
due to confinement-induced valence-band mixing~\cite{Kernreiter2013} and the heavy-hole-type
Rashba spin splitting~\cite{Chesi2007}. The different behavior of confined band electrons
and holes warrants more systematic investigation and, as we will see below, considering
the quasi-1D case sheds new light on the different ramifications of spin-orbit coupling in
interacting systems. Our investigation also reveals that the quasiparticle effective-mass is more strongly suppressed by the
exchange interaction in nanowires with spin splitting.

In addition, quantum wires with strong spin-orbit coupling are
currently attracting great interest as possible hosts of exotic quasiparticle excitations such
as Majorana~\cite{Alicea2012} and fractional~\cite{Klinova2013} fermions. Clarifying the
effect of interactions in such systems is necessary for a complete understanding of experiments
aimed at verifying the existence of the unusual quasi-particle excitations. 

The remainder of this article is organised as follows. We introduce our theoretical model
of a Rashba-spin-split quantum wire in Sec.~\ref{sec:model} and discuss pertinent properties
of the single-particle eigenstates. The formalism for calculating the exchange energy for this
system is presented in Sec.~\ref{sec:exchange}, together with the results. Amongst these is
the ability to express functional dependencies of the exchange energy per particle in terms
of a universal scaling function, and the enhanced suppression of the density-of-states effective
mass. Our findings are summarized, and related to the existing body of knowledge, in
Sec.~\ref{sec:concl}. Certain formal details are given in Appendices.

\section{Theoretical description of Rashba-split nanowire states
\label{sec:model}}

\begin{figure*}
\includegraphics[width=0.32\textwidth]{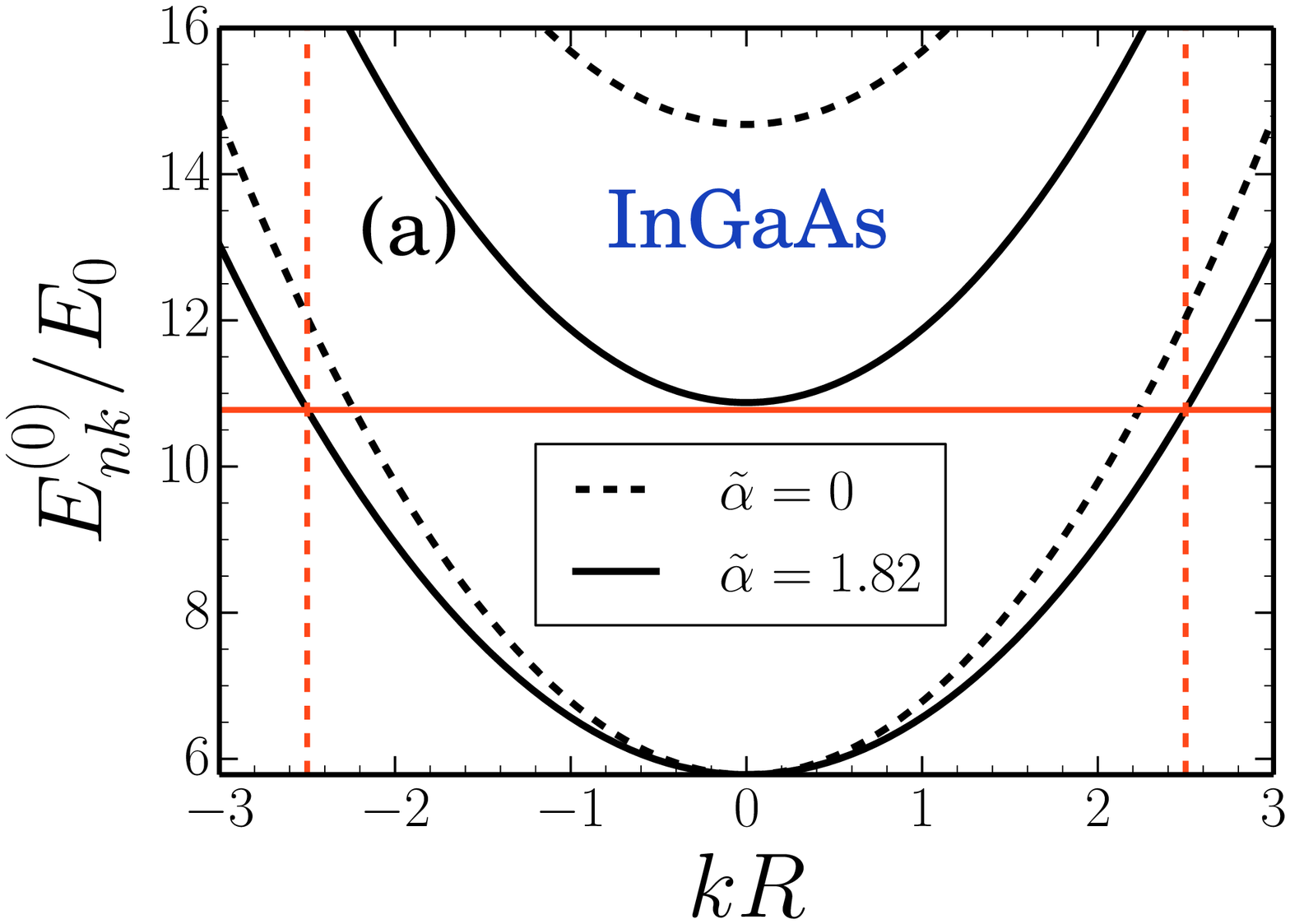}
\hfill\includegraphics[width=0.32\textwidth]{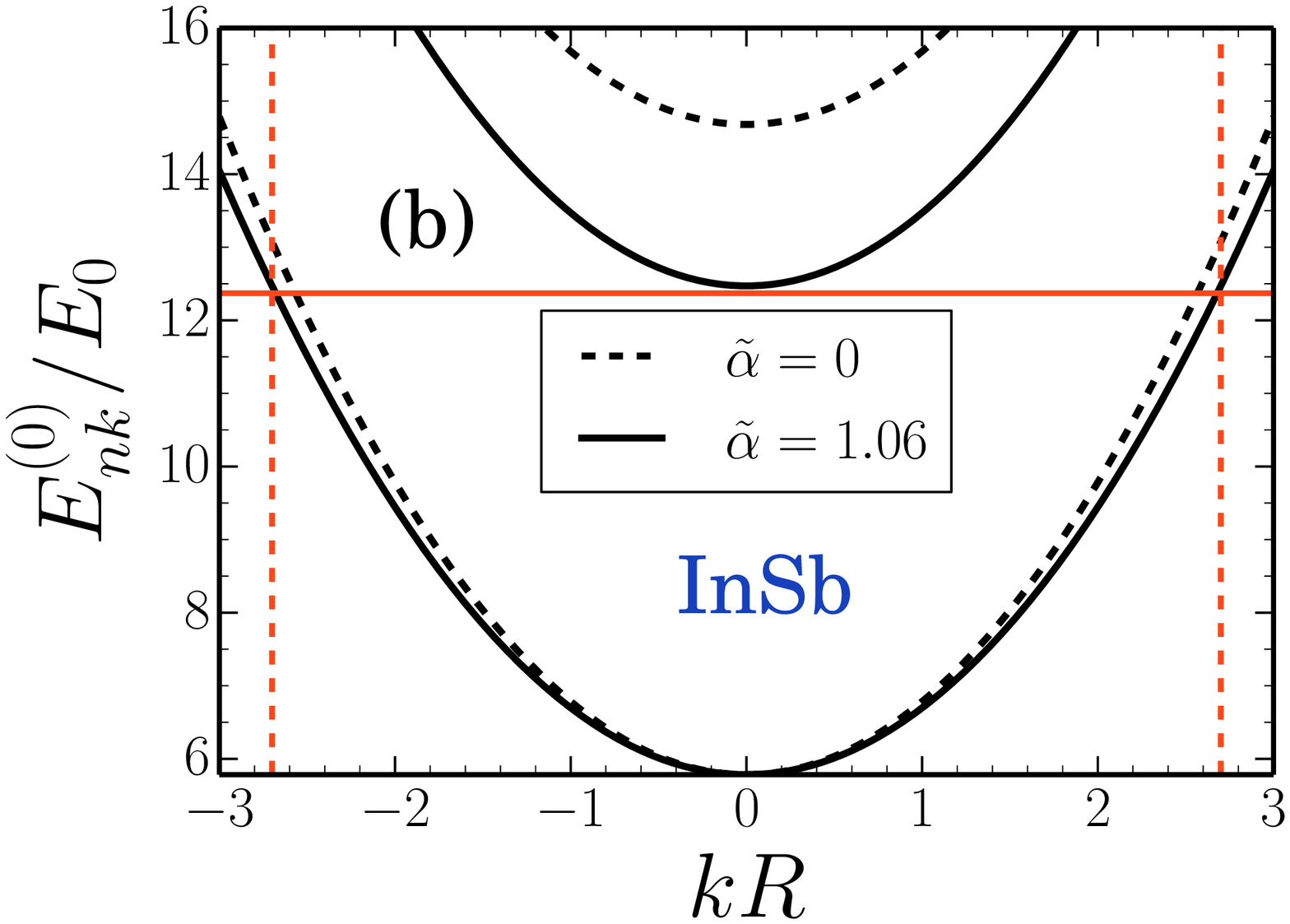}
\hfill\includegraphics[width=0.32\textwidth]{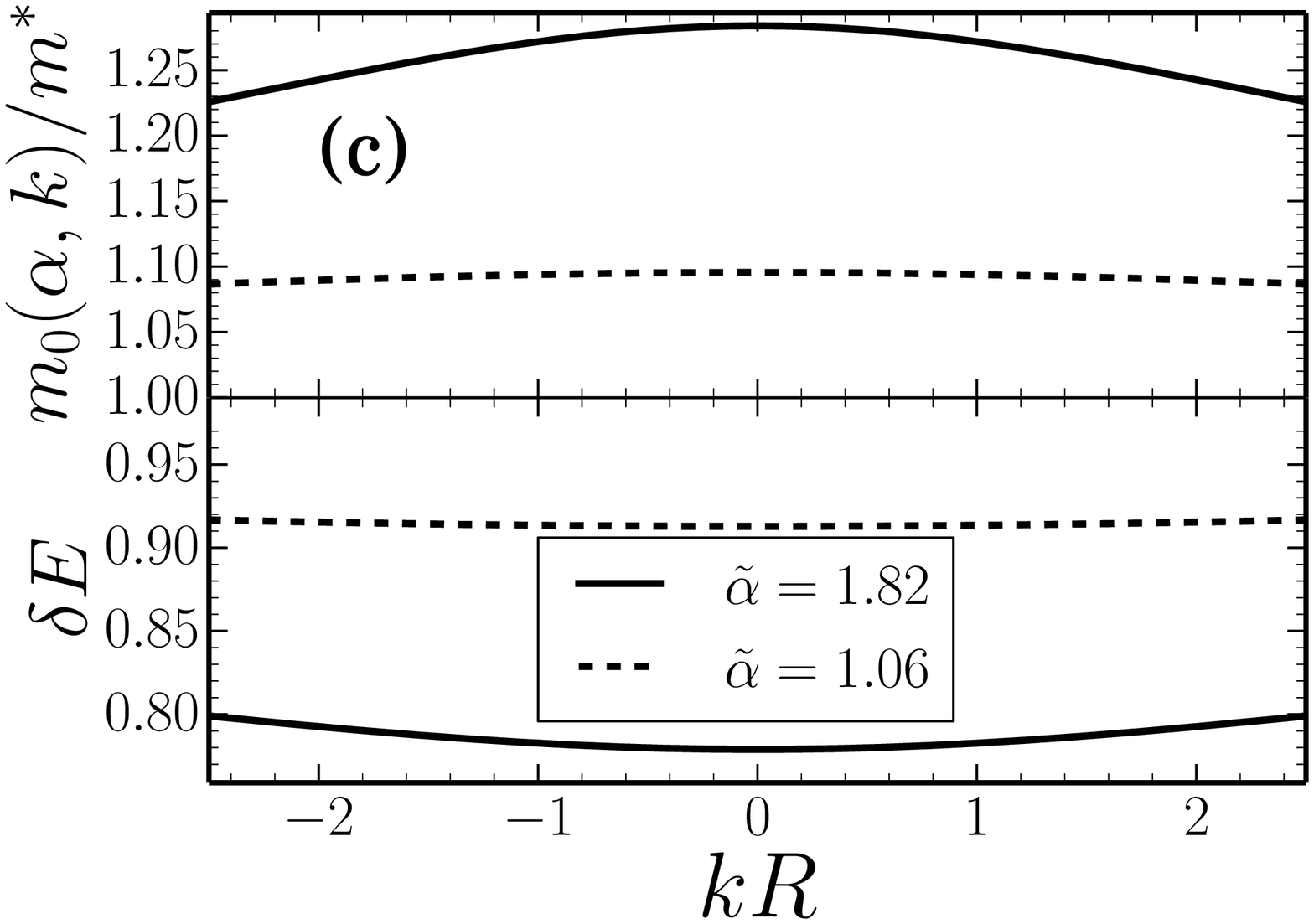}
\caption{Electronic structure of noninteracting electrons in nanowires with
spin splitting induced by a radial electric field. The solid curves in panel~(a) [(b)] show the
single-particle energy dispersions of the lowest two subbands obtained for a value
of $\tilde\alpha$ corresponding to a recent experimental realization using InGaAs [InSb] as
the wire material. To illustrate the effect of spin splitting, the corresponding dispersions for
$\tilde\alpha=0$ are also plotted as dashed curves. Vertical lines are used to indicate the
range of wave numbers for which only the lowest subband is occupied. Panel~(c)
illustrates more quantitatively the effect of spin-orbit coupling on the lowest nanowire-subband
dispersions. In the upper (lower) panel, the ratio of the density-of-states effective masses
(subband energies where $\delta E\equiv [E_k^{(0)}(\alpha)-E_0^{(0)}(\alpha)]/[E_k^{(0)}(0)-E_0^{(0)}(0)]$) for finite and for zero $\tilde\alpha$ are plotted as a function of wave number.}
\label{fig:SOI}
\end{figure*}

In our study, we aim to develop a general understanding of the effect of spin-orbit
coupling on exchange-related many-particle corrections in quasi-1D nanowires. Hence, rather than attempting to
describe the detailed electron density profile for a specific sample based on a self-consistent
Poisson-Schr\"odinger calculation, we consider a model cylindrical quantum wire with radius
$R$ that is defined by a hard-wall potential where a constant radial electric field $\vek{\Ee} =
\Ee\, \vek{\hat r}$ gives rise to a spin-orbit coupling of the Rashba type. In a real sample,
such a radially symmetric field configuration could be generated, e.g., via biasing of
an external gate that is wrapped around the wire surface~\cite{Storm2012}. The
pragmatic assumption of a constant electric-field magnitude is justified in Appendix~\ref{app:EvsR}.
See especially Fig.~\ref{fig:profile}. For our situation of interest, the noninteracting-electron 
dynamics in the wire is described by the Hamiltonian $H = H^{(0)} + U(r)$, where
\begin{equation}
U(r) = \left\{ \begin{array}{cl} 0 & r<R \\ \infty & r\ge R \end{array} \right. \quad ,
\end{equation}
and $H^{(0)}$ is a Rashba-type~\cite{Rashba1960,Bychkov1984} single-electron Hamiltonian 
\begin{equation}\label{eq:HamRashba}
H^{(0)} = \frac{\vek{p}^2}{2m^*}+ \frac{\alpha\, \Ee}{\hbar} \, \vek{\hat{r}}\cdot \left( \vek{\sigma}
\times\vek{p}\right) \quad .
\end{equation}
Here $m^*$ is the band mass of electrons in the semiconductor material making up the
nanowire, $\alpha$ is the material-dependent Rashba spin-orbit-coupling constant, and
$\vek{\sigma}=(\sigma_x,\sigma_y,\sigma_z)^T$ denotes the vector of Pauli
matrices. We will find the confined-electron states in the nanowire by superimposing
solutions of the single-particle Schr\"odinger equation $H^{(0)} \,\psi = E\, \psi$ to satisfy the
cylindrical hard-wall boundary condition.

The Hamiltonian (\ref{eq:HamRashba}) can be conveniently expressed in cylindrical
coordinates $(r,\varphi,z)$ as
\begin{eqnarray}\label{eq:H}
H^{(0)} &=& -\frac{\hbar^2}{2m^*}\left(\frac{\partial^2}{\partial r^2}+ \frac{1}{r}\frac{\partial}
{\partial r}+\frac{1}{r^2}\frac{\partial^2}{\partial \varphi^2}+\frac{\partial^2}{\partial z^2}\right)
\openone \nonumber \\ && +\,\, i\, \alpha \, \Ee \left[ \sigma_z \,\frac{1}{r} \frac{\partial}{\partial 
\varphi} + i \left(\ee^{-i \varphi} \, \sigma_+ -  \ee^{i\varphi}\, \sigma_-  \right)
\frac{\partial}{\partial z}\right] , \quad
\end{eqnarray}
where $\sigma_\pm = (\sigma_x \pm i\,\sigma_y)/2$ are the spin-$1/2$ ladder operators.
The explicit form of (\ref{eq:H}) motivates a separation \textit{Ansatz\/} for the eigenstates
of $H^{(0)}$:
\begin{equation}\label{eq:Ansatz}
\psi(r, \varphi, z) = \frac{\ee^{i k z}}{\sqrt{L}}\,\, \ee^{i \nu \varphi} \,
\ee^{- i \frac{\sigma_z}{2} \varphi} \, \phi_{\nu, k}(r) \quad ,
\end{equation}
where $\phi_{\nu, k}(r)$ is the radial spinor wave function,
 $\nu=\pm1/2, \pm 3/2, \dots$ is an odd half-integer number, $k$ denotes the wave
number associated with the free electron motion in the quantum wire, and $L$ is the wire
length. The resulting radial Schr\"odinger equation that determines $\phi_{\nu, k}(r)$ can be
written in dimensionless form as ${\mathcal H}_{\nu, \kappa}\, \chi_{\nu, \kappa}(\varrho) =
\varepsilon \, \chi_{\nu, \kappa} (\varrho)$, with
\begin{equation}\label{eq:Hr}
{\mathcal H}_{\nu, \kappa} = -\left( \frac{\partial^2}{\partial \varrho^2} + \frac{1}{\varrho}
\frac{\partial}{\partial\varrho}\right) \openone + \frac{\hat m^2}{\varrho^2} -\tilde\alpha\,
\sigma_z \, \frac{\hat m}{\varrho} + \tilde\alpha\, \kappa\, \sigma_y + \kappa^2\,\openone \, ,
\end{equation}
and the definitions $\hat m = \nu\, \openone - \frac{1}{2}\, \sigma_z$, $\varrho = r/R$,
$\kappa = k R$, $\varepsilon = E/E_0$ where $E_0 =\hbar^2/(2m^* R^2)$, $\tilde{\alpha}=
2 R m^* \alpha \Ee/\hbar^2$, and $\phi_{\nu, k}(r)\equiv\chi_{\nu, \kappa}(r/R)/R$.

We employ the subband $\kk\cdot\pp$ method~\cite{bro85,bro85a} to find the
cylindrical-nanowire eigenstates and single-particle subband-energy dispersions $E^{(0)}_{nk}$.
Simultaneous invariance under time reversal ($\sigma_y\, {\mathcal H}^\ast_{-\nu, -\kappa}\,
\sigma_y = {\mathcal H}_{\nu, \kappa}$) and spatial inversion ($\ee^{-i\frac{\pi}{2}\sigma_z}\,
{\mathcal H}_{\nu, -\kappa}\, \ee^{i\frac{\pi}{2}\sigma_z} = {\mathcal H}_{\nu, \kappa}$)
imply that each subband is (at least) doubly degenerate~\cite{Roessler}. The first step is
to find the eigenstates that are associated with the subband-edge energies $E^{(0)}_{n0}$.
These states are then used as a basis set for expressing the eigenstates at general $k \ne 0$;
with expansion coefficients determined from solving a matrix equation that is equivalent to the
Schr\"odinger equation.

The Hamiltonian of Eq.~(\ref{eq:Hr}) is diagonal when $\kappa=0$,
\begin{subequations}
\begin{eqnarray}
{\mathcal H}_{\nu, 0} &=& \left( \begin{array}{cc} {\mathsf H}_\nu & 0 \\ 0 & {\mathsf H}_{-\nu} 
\end{array} \right) \quad , \\ {\mathsf H}_\nu &=& -\left( \frac{\partial^2}{\partial \varrho^2} +
\frac{1}{\varrho} \frac{\partial}{\partial\varrho}\right) + \frac{\left(\nu - \frac{1}{2}\right)^2}
{\varrho^2} - \tilde\alpha\, \frac{\nu - \frac{1}{2}}{\varrho} \, , \quad
\end{eqnarray}
\end{subequations}
hence the subband-edge states are also spin-projection eigenstates of $\sigma_z$ with
eigenvalue $\sigma=\pm 1$. We can therefore write
\begin{subequations}
\begin{equation}
\chi_{\nu, \kappa}(\varrho) = \sum_{n^\prime=1}^\infty \left( c_{\nu, \kappa}^{(n^\prime 
\uparrow)} \ket{\nu, \uparrow, n^\prime} + c_{\nu, \kappa}^{(n^\prime \downarrow)} \ket{\nu,
\downarrow, n^\prime} \right) \, ,
\end{equation}
with the subband-edge basis-state definitions
\begin{eqnarray}
\ket{\nu, \uparrow, n^\prime} &=& \mathcal{F}_{\nu-\frac{1}{2}}^{(\varepsilon_{\nu,
+}^{(n^\prime)})}(\varrho) \left( \begin{array}{c} 1 \\ 0 \end{array} \right) \quad , \\
\ket{\nu, \downarrow, n^\prime} &=& \mathcal{F}_{-\nu-\frac{1}{2}}^{(\varepsilon_{\nu,
-}^{(n^\prime)})}(\varrho) \left( \begin{array}{c} 0 \\ 1 \end{array} \right) \quad ,
\end{eqnarray}
\end{subequations}
and the functions $\mathcal{F}_{\sigma\nu-\frac{1}{2}}^{(\varepsilon_{\nu,
\sigma}^{(n^\prime)})}(\varrho)$ being solutions of the radial-confinement problem defined
by the Hamiltonian ${\mathsf H}_{\sigma \nu} +U(\varrho R)/E_0$ with corresponding
dimensionless eigenenergies $\varepsilon_{\nu, \sigma}^{(n^\prime)}$. 
We number the subband-edge states for fixed $\nu$ and $\sigma$ in ascending order of energy, 
that is $\varepsilon_{\nu, \sigma}^{(n^\prime)}>\varepsilon_{\nu,
\sigma}^{(n^{\prime\prime})}$ when $n^\prime > n^{\prime\prime}$. Time-reversal symmetry
mandates the Kramers degeneracy $\varepsilon_{\nu, \sigma}^{(n^\prime)}=
\varepsilon_{-\nu, -\sigma}^{(n^\prime)}$. See Appendix~\ref{app:radialWF} for more
mathematical details.

The full single-electron subband dispersions can be found from solving the
eigenvalue problem
\begin{subequations}
\begin{widetext}
\begin{equation}\label{eq:kdotp}
\left( \begin{array}{cccccc} \varepsilon_{\nu, +}^{(1)} + \kappa^2 & -i \tilde\alpha \kappa \,
I_\nu^{(11)} & \dots & 0 & -i \tilde\alpha \kappa \, I_\nu^{(1 n^\prime)} & \dots \\ i \tilde\alpha 
\kappa \, \big[I_\nu^{(11)}\big]^\ast & \varepsilon_{\nu, -}^{(1)} + \kappa^2 & \dots & i \tilde
\alpha \kappa \, \big[ I_\nu^{(1 n^\prime)}\big]^\ast & 0 & \dots \\ \vdots & \vdots & \ddots &
\vdots &  \vdots & \ddots \\ 0 &  -i \tilde\alpha \kappa \, I_\nu^{(1 n^\prime)} & \dots &
\varepsilon_{\nu,+}^{(n^\prime)} + \kappa^2 & -i \tilde\alpha \kappa \, I_\nu^{(n^\prime
n^\prime)} & \dots \\ i \tilde\alpha \kappa \, \big[ I_\nu^{(1 n^\prime)} \big]^\ast & 0 & \dots &
i \tilde\alpha \kappa \, \big[ I_\nu^{(n^\prime n^\prime)} \big]^\ast &\varepsilon_{\nu,
-}^{(n^\prime)} + \kappa^2 & \dots \\ \vdots & \vdots & \ddots & \vdots & \vdots & \ddots
\end{array}\right) \left( \begin{array}{c}
c_{\nu, \kappa}^{(1 \uparrow)} \\ c_{\nu, \kappa}^{(1 \downarrow)} \\ \vdots \\ c_{\nu,
\kappa}^{(n^\prime \uparrow)} \\ c_{\nu, \kappa}^{(n^\prime \downarrow)} \\ \vdots \end{array}
\right) = \varepsilon_\nu(\kappa) \left( \begin{array}{c} c_{\nu, \kappa}^{(1 \uparrow)} \\
c_{\nu, \kappa}^{(1 \downarrow)} \\ \vdots \\ c_{\nu, \kappa}^{(n^\prime \uparrow)} \\ c_{\nu,
\kappa}^{(n^\prime \downarrow)} \\ \vdots \end{array}\right) \quad ,
\end{equation}
\end{widetext}
with matrix elements
\begin{equation}
I_\nu^{(n n^\prime)} = 2\pi \int_0^1d\varrho \,\, \varrho\,\, \Big[\mathcal{F}_{\nu-
\frac{1}{2}}^{(\varepsilon_{\nu,+}^{(n)})}(\varrho)\Big]^\ast \,\, \mathcal{F}_{-\nu-
\frac{1}{2}}^{(\varepsilon_{\nu,-}^{(n^\prime)})}(\varrho) \quad .
\end{equation}
\end{subequations}
For the purpose of this study, we only need to obtain the dispersion of the lowest nanowire
subband. We find that, for realistic values of $\tilde\alpha$ (see for instance the examples 
below), truncation of the eigenvalue problem (\ref{eq:kdotp}) to the subspace spanned by
the states $\{ \ket{1/2, \uparrow, 1}, \ket{1/2, \downarrow, 1}\}$ and its time-reversed
counterpart yields sufficiently accurate results. Hence, in the following, we will use the
wave functions
\begin{subequations}\label{eq:lowSub}
\begin{eqnarray}
\psi_1(r, \varphi, z) &=& \frac{\ee^{i k z}}{\sqrt{L}\, R} \big(-i \sin\eta_{k R} \ket{1/2, \uparrow, 1}
\nonumber \\ && \hspace{1.5cm}
+ \,\, \ee^{i\varphi} \cos\eta_{k R} \ket{1/2, \downarrow, 1} \big) , \quad \\
\psi_2(r, \varphi, z) &=& \frac{\ee^{i k z}}{\sqrt{L}\, R} \big(-i \sin\eta_{k R} \ket{-1/2,\downarrow, 1}
\nonumber \\ && \hspace{1cm}
+\,\, \ee^{-i\varphi} \cos\eta_{k R} \ket{-1/2, \uparrow, 1} \big)
\end{eqnarray}
\end{subequations}
to describe lowest-subband states with the dispersion
\begin{eqnarray}\label{eq:energylowsubband}
E^{(0)}_{1k} \equiv E^{(0)}_{2k} &=& E_0 \left[ (k R)^2+\frac{1}{2}\left(\varepsilon_{1/2, +}^{(1)} +
\varepsilon_{1/2,-}^{(1)}\right) \right.\nonumber\\[2mm] {}&& \hspace{-1.5cm} - \left.
\frac{1}{2}\sqrt{\left( \varepsilon_{1/2, +}^{(1)} - \varepsilon_{1/2,-}^{(1)} \right)^2 + \left( 2
\tilde \alpha k R\, I_{1/2}^{(11)}\right)^2} \right] . \quad
\end{eqnarray}
The coefficients entering Eqs.~(\ref{eq:lowSub}) are
\begin{subequations}
\label{eq:lowcoeff}
\begin{eqnarray}
\sin\eta_\kappa &=& \nonumber \\ && \hspace{-0.5cm} \frac{1}{\sqrt{2}}\left(1+\frac{\left|
\varepsilon_{1/2, +}^{(1)} - \varepsilon_{1/2,-}^{(1)}\right|}{\sqrt{\left( \varepsilon_{1/2, +}^{(1)} 
- \varepsilon_{1/2,-}^{(1)} \right)^2 + \left( 2 \tilde \alpha \kappa I_{1/2}^{(11)}\right)^2}}
\right)^{\frac{1}{2}} , \nonumber \\[2mm] \\[2mm]
\cos\eta_\kappa &=& \nonumber \\ && \hspace{-0.5cm} \frac{1}{\sqrt{2}}\left(1-\frac{\left|
\varepsilon_{1/2, +}^{(1)} - \varepsilon_{1/2, -}^{(1)}\right|}{\sqrt{\left( \varepsilon_{1/2, +}^{(1)}
- \varepsilon_{1/2,-}^{(1)} \right)^2 + \left( 2 \tilde\alpha \kappa I_{1/2}^{(11)}\right)^2}}
\right)^{\frac{1}{2}} , \nonumber \\[2mm]
\end{eqnarray}
\end{subequations}

\begin{figure}[t]
\includegraphics[width=0.7\columnwidth]{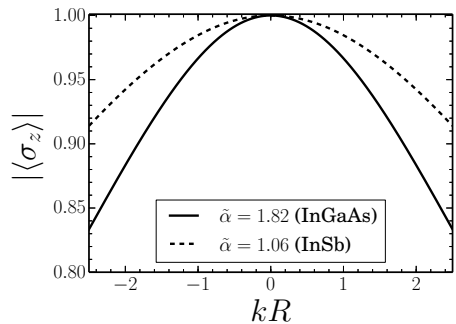}
\caption{\label{fig:Spinz}
The magnitude of the expectation value for spin
projection parallel to the wire axis, $|\av{\sigma_z}| = \av{\sigma_z}_1 = - \av{\sigma_z}_2
\equiv 2\, \sin^2\eta_{k R}-1$, for states from the lowest doubly degenerate ($n=1$ and
$2$) subband.} 
\end{figure}

\begin{table}[b]
\caption{\label{tab:States}
Properties of the three lowest doubly degenerate nanowire subband edges obtained for
parameters applicable to recent experimental realizations.}
\tabcolsep 1ex
\renewcommand{\arraystretch}{1.2}
\begin{tabular}{ccccc}
\hline\hline
subband  & $E^{(0)}_{n0}/E_0$ for & $E^{(0)}_{n0}/E_0$ for & subband-edge\\[-0.1cm]
index $n$ &  $\tilde\alpha = 1.82$ & $\tilde\alpha  = 1.06$ & (basis) state \\ \hline
1 & 5.783 &5.783 & $\ket{+\frac{1}{2}, \uparrow,1}$\\ 
2 & 5.783 &5.783 & $\ket{-\frac{1}{2}, \downarrow,1}$\\ 
3 &10.87 &12.47 &  $\ket{+\frac{3}{2},\uparrow,1}$ \\
4 &10.87 &12.47 &  $\ket{-\frac{3}{2},\downarrow,1}$ \\
5 &18.35 &16.85 & $\ket{-\frac{1}{2},\uparrow,1}$ \\
6 &18.35 &16.85 & $\ket{+\frac{1}{2},\downarrow,1}$ \\
\hline\hline
\end{tabular}
\end{table}

Figure~\ref{fig:SOI} illustrates the noninteracting-electron band structure of nanowires using
parameters relevant to  recent experimental realizations\footnote{Even
though the type of Rashba spin splitting realized in these experiments differs from that
adopted in our work, the same bandstructure-related prefactors and orders of magnitude
for the electric field apply in both their and our systems. This motivates our use of
spin-orbit-coupling strengths measured in Refs.~\onlinecite{Schaepers2004,VanWeperen2014}.}
in Ref.~\onlinecite{Schaepers2004} (InGaAs material with conduction-band effective mass
$m^*=0.037\, m_0$, where $m_0$ is the electron mass in vacuum, $R=300\,\mathrm{nm}$,
and $\alpha\, \Ee = 10^{-11}\,\mathrm{eV\,m}$), and Ref.~\onlinecite{VanWeperen2014}
(InSb, $m^*=0.013\, m_0$, $R=50\,\mathrm{nm}$, $\alpha\, \Ee=10^{-10}\,\mathrm{eV\,m}$).
Within our model, the relevant quantity determining the effect of spin-orbit coupling is
$\tilde\alpha$, which is equal to $1.82$ and $1.06$ for the InGaAs and InSb nanowires,
respectively. For comparison, we show also the result for $\tilde\alpha=0$. As the lowest
subband-edge states have quantum numbers $\{\nu=1/2, \uparrow\}$ and $\{\nu=-1/2,
\downarrow\}$, respectively, their energy is independent of $\tilde\alpha$, and the spin-orbit
coupling only affects the dispersion at finite $k$. In panel~(c), the upper plot shows
the ratio of the single-particle density-of-states effective mass,
\begin{equation}
m_0(\alpha,k) = \frac{\hbar^2 k}{\partial E^{(0)}_{1k}/\partial k}\quad ,
\end{equation}
of the lowest subband with and without spin-orbit coupling for the two values of
$\tilde\alpha$. The lower panel in panel~(c) illustrates the relative change in energy for the
lowest-subband states due to the spin-orbit coupling. As can be seen from the plot, the
renormalization of the single-particle effective mass due to spin-orbit coupling can
amount to up to 30\% (for $\tilde\alpha=1.82$) and also depends appreciably on the value of
the wave vector. In Fig.~\ref{fig:Spinz}, we show the magnitude of the expectation value of the
spin projection along the wire axis for the lowest subband  as a function of  the wave number
$k$. It decreases with increasing $k$, as the states  given in Eqs.~(\ref{eq:lowSub}) together
with  (\ref{eq:lowcoeff}) become superpositions of $\uparrow$ and $\downarrow$  states for
finite $k$. Table~\ref{tab:States} summarizes properties of the three lowest doubly degenerate
subband edges in the two material systems. 
Note the rather large energy splitting of the (doubly degenerate) next-to-lowest subbbands due to the spin-orbit coupling.
Without spin-orbit coupling ($\tilde\alpha=0$) the band edge energy of the subbands $n=3,\dots,6$ is $E^{(0)}_{n0}/E_0\approx 14.68$.

\section{Effect of spin-orbit coupling on the Coulomb-exchange energy
\label{sec:exchange}}

The Coulomb exchange interaction between electrons renormalizes the
quasiparticle dispersion of nanowire subbands, which is then given by~\cite{Giuliani2005}
\begin{eqnarray}
E^{(\text{int})}_{nk}=E^{(0)}_{nk} + \Sigma^{(\text{X})}_{nk}
\end{eqnarray}
in terms of the non-interacting subband energy dispersion $E^{(0)}_{nk}$ obtained in the
previous Section and the exchange (Fock) self-energy
\begin{eqnarray}
\Sigma^{(\text{X})}_{nk} = -\sum_{n'}\int \frac{dk'}{2\pi}~V^{(n n')}_{k k'} \, n_{\text{F}}(E_{n'k'})~.
\end{eqnarray}
Here $n_{\text{F}}(E)$ denotes the Fermi-Dirac distribution function, and $V^{(n n')}_{k k'}$ is
the matrix element of Coulomb
interaction between nanowire-electron states given by
\begin{align}\label{eq:Coulomb}
V^{(nn')}_{kk'}&=C\int d^2 r_\perp \int d^2r'_\perp \int_{-L/2}^{L/2} dz \, \frac{\ee^{i ( k^\prime
- k) z}}{\sqrt{z^2+\big|\vek{r_\perp}-\vek{r^\prime_\perp}\big|^2}}\nonumber\\[0.2cm]
&\hspace{0.8cm}\times\, \xi^\dagger_{n'k'}(\vek{r_\perp})\, \xi_{nk}(\vek{r_\perp})\,\,\,\,
\xi^\dagger_{nk}(\vek{r^\prime_\perp})\, \xi_{n'k'}(\vek{r^\prime_\perp})~,
\end{align}
where $C\equiv e^2/(4\pi\varepsilon_0\varepsilon_{\mathrm r})$ is the Coulomb-interaction
strength, $\vek{r_\perp}\equiv(r,\varphi)$ denotes the position vector in the coordinates
perpendicular to the wire axis, and $\xi_{nk}(\vek{r_\perp})\equiv \ee^{i \nu \varphi}\,\ee^{- i
\frac{\sigma_z}{2} \varphi} \, \phi_{\nu, k}(r)$ is the transverse spinor part of the wavefunction
in Eq.~(\ref{eq:Ansatz}). In the following, we consider the zero-temperature
limit and thus replace the Fermi-Dirac distribution function by
$n_{\text{F}}(E)\equiv\Theta(E_{\text{F}}-E)$, with $\Theta(E)$ being
the Heaviside step function and $E_{\text{F}}$ denoting the Fermi energy. The condition
$E_{nk} \equiv E_{\text{F}}$ defines the Fermi wave vectors $k_{\text{F}n}$ for occupied
nanowire subbands.
We now focus on the low-density situation where only states in the lowest doubly degenerate
subband are occupied up to the Fermi wave vector $k_{\text{F}}=k_{\text{F}1} \equiv
k_{\text{F}2}$. For this situation, we can write
\begin{subequations}
\begin{eqnarray}
\Sigma^{(\text{X})}_{nk}=\frac{-2C}{R}
\left[
\tilde\Lambda_{\text{intra}}(\tilde\alpha,\kappa_{\text{F}},\kappa)+
\tilde\Lambda_{\text{inter}}(\tilde\alpha,\kappa_{\text{F}},\kappa)
\right]~,\nonumber\\
\end{eqnarray}
where $\tilde\Lambda_\mathrm{intra}$ ($\tilde\Lambda_\mathrm{inter}$) includes contributions 
arising from the exchange interaction between particles from the same band (from different
bands). In the limit $L\to\infty$, we obtain the explicit expressions
\begin{widetext}
\begin{eqnarray}\label{eq:intraEx}
\tilde\Lambda_\mathrm{intra}(\tilde\alpha, \kappa_\mathrm{F},\kappa) &=& 
\int_{-\kappa_{\text{F}}}^{\kappa_{\text{F}}} d\kappa' ~~\int_0^1 d\varrho\, \varrho\, \int_0^1
d\varrho' \,
\varrho' \int_0^{2\pi} d\tilde\varphi~~K_0\left(|\kappa - \kappa' | \sqrt{\varrho^2+\varrho'^2
- 2 \varrho \varrho' \, \cos\tilde\varphi}\right) \nonumber \\[2mm] && \hspace{-1cm} \times
\left[ \sin^2\eta_\kappa\,  \sin^2\eta_{\kappa'} \Big| \mathcal{F}_0^{(\varepsilon_{1/2,+}^{(1)})}
(\varrho) \Big|^2 \, \Big| \mathcal{F}_0^{(\varepsilon_{1/2,+}^{(1)})}(\varrho')\Big|^2 +
\cos^2\eta_\kappa\, \cos^2\eta_{\kappa'} \Big| \mathcal{F}_{-1}^{(\varepsilon_{1/2,-}^{(1)})}
(\varrho) \Big|^2 \Big| \mathcal{F}_{-1}^{(\varepsilon_{1/2,-}^{(1)})}(\varrho') \Big|^2 \right. 
\nonumber\\[2mm] &&\hspace{-1cm} + \, \left. \sin\alpha_\kappa\, \cos\alpha_\kappa\, \sin
\alpha_{\kappa'}\, \cos\alpha_{\kappa'}\Big( \Big| \mathcal{F}_0^{(\varepsilon_{1/2,+}^{(1)})}
(\varrho) \Big|^2 \, \Big| \mathcal{F}_{-1}^{(\varepsilon_{1/2,-}^{(1)})}(\varrho') \Big|^2 + \Big|
\mathcal{F}_0^{(\varepsilon_{1/2,+}^{(1)})}(\varrho') \Big|^2 \, \Big|
\mathcal{F}_{-1}^{(\varepsilon_{1/2,-}^{(1)})}(\varrho) \Big|^2\Big)\right] , \quad \\
\tilde\Lambda_\mathrm{inter}(\tilde\alpha, \kappa_\mathrm{F},\kappa) &=&
\int_{-\kappa_{\text{F}}}^{\kappa_{\text{F}}} d\kappa' ~ \sin^2(\eta_\kappa-\eta_{\kappa'})
~\int_0^1 d\varrho\, \varrho\, \int_0^1 d\varrho' \, \varrho' \int_0^{2\pi} d\tilde\varphi~\cos
\tilde \varphi \nonumber \\[2mm] && \hspace{-0.5cm} \times\,\, K_0\left(|\kappa - \kappa' |
\sqrt{\varrho^2+\varrho'^2 - 2 \varrho \varrho'\, \cos\tilde\varphi}\right)~
\mathcal{F}_0^{(\varepsilon_{1/2,+}^{(1)})}(\varrho)~~
\mathcal{F}_{-1}^{(\varepsilon_{1/2,-}^{(1)})}(\varrho)~~
\mathcal{F}_0^{(\varepsilon_{1/2,+}^{(1)})}(\varrho')~~
\mathcal{F}_{-1}^{(\varepsilon_{1/2,-}^{(1)})}(\varrho')~,
\end{eqnarray}
\end{widetext}
\end{subequations}
where $K_0$ is the modified Bessel function of the second kind\cite{Abramowitz}.
For the numerical evaluation of the intra-band contribution (\ref{eq:intraEx}), we employ a
modified quadrature method~\cite{Chao1991}, described in greater detail in
Appendix~\ref{app:MQM}, to deal with the logarithmic singularity encountered when the
argument of $K_0(\cdot)$ approaches zero.

\begin{figure}[b]
\includegraphics[width=0.7\columnwidth]{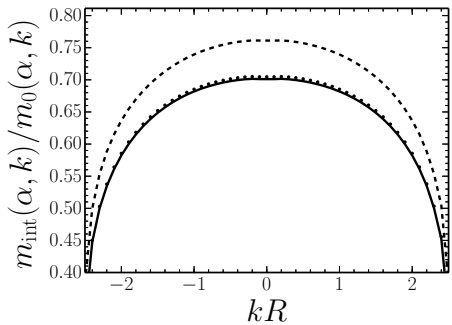}
\caption{\label{fig:IntEffmass}
The ratio of the exchange-renormalized effective quasiparticle mass $m_{\text{int}}(\alpha,k)$
to the bare single-particle effective mass $m_{0}(\alpha,k)$ is plotted as a function of wave
vector $k$ for states from the lowest nanowire subband assuming a material with dielectric
constant $\varepsilon_\mathrm{r} = 12.9$, $k_{\text{F}}R=2.5$, and spin-orbit-coupling
strength $\tilde\alpha=1.82$ ($\tilde\alpha=0$) as the solid (dashed) curve.
For comparison, the dotted curve shows the result obtained under the assumption
that the electric field strength varies linearly with the radial coordinate near the wire's center, as
described in Appendix \ref{app:EvsR}.} 
\end{figure}

The exchange-renormalized density-of-states (quasiparticle) effective mass for the lowest
subband can be calculated from
\begin{equation}
m_{\text{int}}(\alpha,k)=\frac{\hbar^2 k}{\frac{\partial E_{1k}^{(0)}}{\partial k} +
\frac{\partial \Sigma_{1k}^{(0)}}{\partial k}} \quad .
\end{equation}
In Fig.~\ref{fig:IntEffmass}, we compare the suppression of the quasiparticle effective mass
due to the exchange interaction in a nanowire with finite spin-orbit coupling with that of an
identical nanowire having zero spin-orbit coupling. As can be seen, the presence of spin-orbit
coupling further suppresses the exchange-related quasiparticle-mass by 10-15\% for
the parameters used in our calculation.
Note also the strong wave-vector dependence of the exchange-renormalized quasiparticle effective mass.

\begin{figure}[b]
\includegraphics[width=0.7\columnwidth]{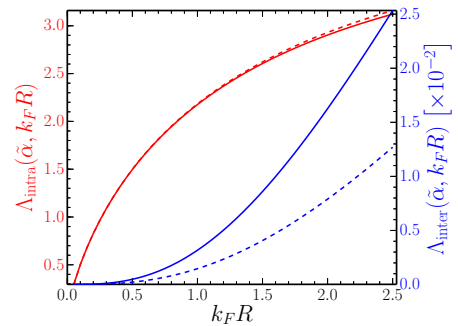}
\caption{\label{fig:Ex}
Scaling functions $\Lambda_\mathrm{intra}$ and $\Lambda_\mathrm{inter}$ associated
with the intra-band and inter-band contributions to the exchange energy per particle in 
cylindrical nanowires with spin-orbit coupling (note the scale of $10^{-2}$ for the inter-band 
contribution). Dashed (solid) curves corresponds to $\tilde\alpha = 1.06$ ($1.82$).} 
\end{figure}

The total exchange energy per particle for the nanowire-electron system is given
by~\cite{Giuliani2005}
\begin{equation}\label{eq:Ex-3}
\frac{E_{\text{X}}}{N} = \frac{1}{2\rho}\sum_{n}\int\frac{dk}{2\pi} \,\,
\Sigma_{n k}^{(\text{X})}\,\, n_{\text{F}}(E_{nk})~,
\end{equation}
where $\rho=N/L$ is the quasi-1D electron density. Again we focus on the low-density
situation where only states in the lowest doubly degenerate subband are occupied up to the
Fermi wave vector. For this situation, we can write
\begin{equation}\label{eq:Ex-4}
\frac{E_{\text{X}}}{N}=-\frac{C}{2 R} \, \left[ \Lambda_\mathrm{intra}(\tilde\alpha,
\kappa_{\text{F}}) + \Lambda_\mathrm{inter}(\tilde\alpha, \kappa_{\text{F}}) \right]~,
\end{equation}
where  
$\Lambda_\mathrm{intra}(\tilde\alpha,\kappa_{\text{F}})=\kappa_\mathrm{F}^{-1}
\int_{-\kappa_{\text{F}}}^{\kappa_{\text{F}}} d\kappa~\tilde\Lambda_\mathrm{intra}(\tilde
\alpha,\kappa_{\text{F}},\kappa)$, and the analogous expression applies for
$\Lambda_\mathrm{inter}$.
Figure~\ref{fig:Ex} illustrates the functional
dependences and relative magnitudes of $\Lambda_\mathrm{intra}$ and
$\Lambda_\mathrm{inter}$. As can be seen, the intra-band contribution is generally
dominant and weakly dependent on $\tilde\alpha$ values considered here. In contrast, the
inter-band contribution changes significantly as a function of $\tilde\alpha$.

\begin{figure}[t]
\includegraphics[width=0.7\columnwidth]{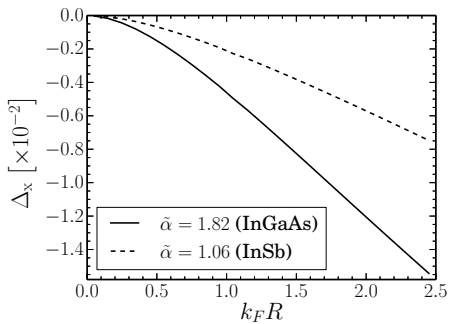}
\caption{\label{fig:Exdiff}
Relative change $\Delta_\mathrm{X}$ in the magnitude of the exchange energy resulting
from a finite Rashba-type spin-orbit coupling quantified by parameter $\tilde\alpha$, as
defined in Eq.~(\ref{eq:ExDiff}). Note the scale factor of $10^{-2}$ for the abscissa. The
dashed (solid) curve shows the result obtained for $\tilde\alpha = 1.06$ ($1.82$), which
corresponds to a recent experimental realization using InSb (InGaAs) as the wire material.} 
\end{figure}

For quantum wires without spin splitting, i.e., in the case $\tilde\alpha=0$, the exchange
energy per particle was found to obey a universal scaling
form~\cite{Gold1990,Calmels1995,Calmels1997}. Our expression for $E_\mathrm{X}/N$
given in Eq.~(\ref{eq:Ex-4}) generalizes these previous results to the case where spin-orbit
coupling is finite. The change in magnitude of the exchange energy arising from finite
$\tilde\alpha$ can be quantified through the relative difference 
\begin{equation}\label{eq:ExDiff}
\Delta_\mathrm{X}=\frac{E_\mathrm{X}(\tilde{\alpha}\neq0)}{E_\mathrm{X}(\tilde{\alpha}=0)}
- 1 \quad ,
\end{equation}
which is visualized in Fig.~\ref{fig:Exdiff}. For the values of $\tilde\alpha$ that correspond
to recent experimental realizations using  InGaAs~\cite{Schaepers2004} and
InSb~\cite{VanWeperen2014}, the associated change amounts to
a \emph{suppression\/} of the exchange-energy magnitude which can be up to 1.6\%. This
behavior is markedly different from the case of a 2D electron system where Rashba spin
splitting has been shown~\cite{Chesi2011} to result in an \emph{increase\/} of the exchange
energy that is roughly one order of magnitude smaller. Thus the Rashba-type spin-orbit coupling due to a radial electric field 
in a cylindrical nanowire system is more similar to a 2D hole system where the interplay between quantum confinement
and spin-orbit effects also results in a suppression of the exchange energy~\cite{Kernreiter2013}.

\section{Conclusions\label{sec:concl}}

We have studied theoretically the electronic properties of the quasi-1D electron system
realized in a cylindrical quantum wire subject to a radially symmetric Rashba-type spin-orbit coupling. 
We determined the single-particle states for a hard-wall confinement using
subband $\kk\cdot \pp$ theory. Focusing on the situation where only the lowest quasi-1D
subband is occupied, we observed that the corresponding energy dispersion can be very 
accurately (to within 0.5\% error) calculated from an effective 2$\times$2 Hamiltonian. 
Taking the material parameters of two experimentally studied nanowire systems (one based
on InGaAs and the other on InSb) as input, we have determined the influence of the
spin-orbit strength on the lowest quasi-1D subband's energy dispersion and on the spin
projection of its corresponding eigenstates parallel to the wire axis, finding both quantities
to be affected by tens of percent due to the presence of spin-orbit coupling.
In particular, the density-of-states effective mass of the noninteracting system turns
out to be increased by 20-25\% for parameters applicable to the InSb nanowires.

With single-particle states in hand, we calculated the quasiparticle effective mass for
the lowest subband and found its exchange-related suppression to be significantly larger in magnitude
(by 10-15\% for parameters used in our calculations) when spin-orbit coupling is finite. In contrast, the
magnitude of the exchange energy per particle is marginally reduced (by upto 1.6\%) by spin-orbit
coupling effects. Thus we find that any meaningful discussion of the interplay between spin-orbit
coupling and exchange interactions in quantum wires needs to be carefully focused on specific
physical quantities, as their relevant parametric dependences can be quite different, both qualitatively
and quantitatively. Furthermore, often the relevance of interaction effects in an electron system
is quantified in relative terms by a parameter $r_\mathrm{s}$ that is related to the
ratio of contributions to the total energy arising from interactions and the single-electron dispersion,
respectively~\cite{Giuliani2005}. In the present context, spin splitting causes an
increase in the single-particle effective mass of quasi-1D electrons simultaneously with the
suppression of the exchange energy. As the relative change in the increase in noninteracting
system's effective mass is an order of magnitude larger than the relative decrease of the
exchange energy, the relative importance of interactions as measured by $r_\mathrm{s}$
turns out to be enhanced by spin-orbit coupling.~\footnote{S. Chesi, private communication.}

While we have focused on a specific configuration of confinement and spin-orbit coupling,
our general results and overall conclusions can be expected to apply also to other
spin-orbit-coupled nanowire systems, e.g., the one considered in Ref.~\onlinecite{Bringer2004}.

\begin{acknowledgments}

We gratefully acknowledge useful discussions with S.~Chesi.

\end{acknowledgments}

\appendix

\section{Radial electric-field profile\label{app:EvsR}}

A proper self-consistent treatment of electrostatic effects generally requires the application
of an iterative Schr\"odinger-Poisson solver method that is specifically adapted to the sample
layout. An added complication arises from the intricate way how the Rashba spin-orbit coupling
strength needs to be determined from expectation values of the electric field taken in a multi-band
bound state~\cite{Winkler2003}. As we intend to focus on the broad implications of spin-orbit
coupling in confined systems, we decided to make an assumption about the radial profile of the
electric field entering in the spin-orbit term that enables us to obtain rather general physical insights.
Here we show the basic consistency of this assumption with the electrostatics of the bound-state
configuration for our system.

Application of Gauss's law using the cylindrical symmetry of the nanowire geometry
yields the relation
\begin{widetext}
\begin{equation}
2\pi r\, L\, \Ee(r) = \frac{-e}{\varepsilon_0\varepsilon_\mathrm{r}} \sum_j
\sum_{|k| \le k_\mathrm{F}} \int_0^L \! dz \int_0^{2\pi} \! d\varphi \int_0^r \!
dr' \, r' \,\, \left[ \psi_j(r', \varphi, z)\right]^\dagger \psi_j(r', \varphi, z) \quad ,
\end{equation}
\end{widetext}
with the single-particle wave functions $\psi_{1,2}(r, \varphi, z)$ given in
Eq.~(\ref{eq:lowSub}). Straightforward calculation yields $\Ee(r) = \Ee_0
\left[ S_{k_\mathrm{F} R} \,\, \mathcal{P}_0(r/R) + C_{k_\mathrm{F} R} \,\,
\mathcal{P}_1(r/R)\right]$, where $\Ee_0 = -N e/(2\pi R L \, \varepsilon_0
\varepsilon_\mathrm{r})$ is an overall scale containing the number of particles
$N$, $S_\kappa = \frac{1}{\kappa} \int_0^{\kappa} d\kappa' \, \sin^2 \eta_{\kappa'}$
and $C_\kappa = \frac{1}{\kappa} \int_0^{\kappa}d\kappa' \, \cos^2 \eta_{\kappa'}$
are weightings of the mixed bound-state contributions for the lowest nanowire
subband, and
\begin{equation}
\mathcal{P}_{0(1)}(\varrho) =  \frac{2\pi}{\varrho} \int_0^{\varrho} \!
d\varrho' \, \varrho' \,\, \left[\mathcal{F}_{0(1)}^{(\varepsilon_{1/2,+(-)}^{(1)})}
(\varrho^\prime)\right]^2
\end{equation}
are the radial density profiles associated with the relevant bound states.

\begin{figure}[b]
\includegraphics[width=7cm]{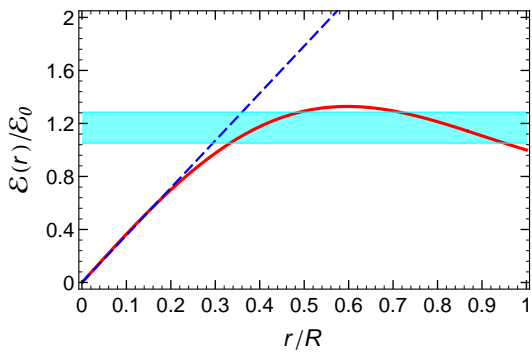}
\caption{\label{fig:profile}
The radial profile of the electric-field magnitude $\Ee(r)$ associated with our calculated
nanowire states is plotted as the solid red curve. The asymptotically linear behaviour for
small $r/R$ is captured by the blue dashed line. The horizontal (cyan) band indicates
the range of field values that lie within 10\% of the mean value. The weak variation of
$\Ee(r)$ for $r\gtrsim 0.3\, R$ motivates our assumption of a constant field magnitude
for the radially symmetric Rashba term in Eq.~(\ref{eq:HamRashba}). The scale of the electric field
is $e\Ee_0=1.6 \kappa_{\text{F}}~\mu\text{eV}\text{nm}^{-1}$ for the scenario based on InGaAs 
while it is $e\Ee_0=41 \kappa_{\text{F}}~\mu\text{eV}\text{nm}^{-1}$ for InSb.} 
\end{figure}

The calculated full electric-field profile is shown in Fig.~\ref{fig:profile}. Our results from
the main paper suggest that generally $S_{k_\mathrm{F} R}\approx 1$, $C_{k_\mathrm{F}
R}\approx 0$; hence $\Ee(r)$ should be essentially determined by the $\mathcal{P}_0(r/R)$
contribution. This is indeed observed in the numerical evaluation. Also, as expected from
the shape of the density profile associated with the $m=0$ bound-state wave function (cf.\
Appendix~\ref{app:radialWF}), the leading behaviour at $r/R\ll 1$ is linear. However, over
most of the wire's cross-section, the field profile is quite well approximated by a constant,
which supports our pragmatic assumption. It is also observed from direct calculation that
$S_{k_\mathrm{F} R}$ and $C_{k_\mathrm{F} R}$ are almost constant in the relevant range
$k_\mathrm{F} R< 2.5$ where only the lowest nanowire subband is occupied.

\begin{figure}[t]
\includegraphics[width=6.8cm]{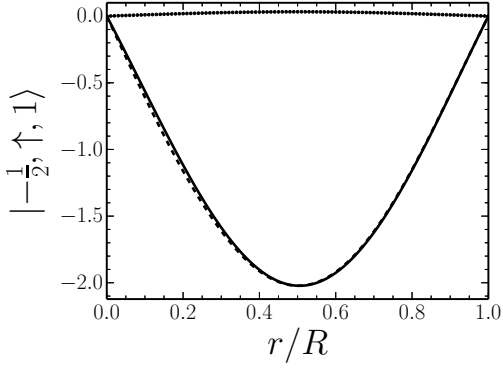}
\caption{\label{fig:secondexstate} 
Real part (dashed curve) and imaginary part (dotted curve) of the subband 
edge wave function for spin up of the second-excited state obtained 
for an electric field that depends linearly on the wire's radius upto $\varrho=0.3$ 
and is constant for $\varrho>0.3$. For comparison we show the real wave function of the corresponding state (solid curve) obtained 
under the assumption of a throughout constant electric field.}
\end{figure}

%%%%%%%%%%%%%%%%%%%%%%%%%%%%%%%%%%%%%%%%%%%%%%%%%
In order to confirm that the omission of the linear electric field dependence for
$r/R\lsim 0.3$ will not alter our conclusions, we consider an electric field which is modelled by
a linear dependence on the radial coordinate upto $\varrho < \varrho_0$ and a constant for
$\varrho>\varrho_0$. In terms of the dimensionless Hamiltonian description in Eq.~(\ref{eq:Hr}),
this implies the replacement $\tilde\alpha\to \tilde\alpha \left[\frac{\varrho}{\varrho_0}\Theta
(\varrho_0-\varrho)+\Theta(\varrho-\varrho_0)\right]$. 
Proceeding as in the case of a constant electric field, we find for the wave functions in the region
with $\varrho<\varrho_0$ the Bessel-function solutions $J_m(\varrho\sqrt{\varepsilon+m\tilde
\alpha/\varrho_0})$ . For the region $\varrho>\varrho_0$, we obtain wave functions which are a
superposition of modified Laguerre functions (see Appendix~\ref{app:radialWF}) and confluent
hypergeometric functions of the second kind. Applying the standard matching conditions at
$\varrho=\varrho_0$ to ensure continuity of the wave functions and their products with the
velocity-operator in the transverse direction determines the unknown coefficients. In this context,
it should be noted that the lowest state is independent of the electric field. Considering the scenario
with $\tilde\alpha=1.82$ as an example and taking into account the hard-wall boundary condition,
we find only small changes for the band-edge energies $E^{(0)}_{(3,4)0}/E_0=11.07$ and
$E^{(0)}_{(5,6)0}/E_0=18.22$ when $\varrho_0=0.3$ (cf. Table~\ref{tab:States}). In
Fig.~\ref{fig:secondexstate} we show the real part (dashed curve) and imaginary part (dotted curve)
of the subband 
edge wave function for spin up of the second-excited state $|-1/2,\uparrow,1\rangle$ and compare this with the corresponding
wave function obtained under the assumption of a constant radial electric-field strength. We
can therefore conclude that the linear electric field dependence for $\varrho<\varrho_0$ changes
the relevant wave functions used in our calculations only slightly. For $\kappa\neq 0$, we find that
the matrix element $I_{1/2}^{(11)}\approx -0.884$, while it is $I_{1/2}^{(11)}\approx -0.916$ with
the assumption of a constant electric field, yielding only sub-percent changes for the dispersions
and exchange-related quantities (see for instance Fig.~\ref{fig:IntEffmass}).

\section{Solution of the radial-confinement problem\label{app:radialWF}}

The general solution of the differential equations present in the diagonal entries of
Eq.~(\ref{eq:Hr}) are power series, given by,
\begin{equation}\label{eq:genPoly}
\mathcal{F}_{m}^{(\varepsilon_{\nu,\pm}^{(n^\prime)})}(\varrho)=\varrho^m(a_0+a_1
\varrho+\sum_{n=2}^{\infty}a_n\varrho^n)
\end{equation}
which fulfill the relation $\mathcal{F}_{-m}^{(\varepsilon_{\nu,\pm}^{(n^\prime)})}(\varrho)=
\mathcal{F}_{+m}^{(\varepsilon_{\nu,\mp}^{(n^\prime)})}(\varrho)$ yielding the eigenstates.
Disregarding the ill-behaved and unphysical part in the expansion at the origin,  
the coefficients of the polynomials are determined by the recursion relation
\begin{align}\label{eq:polynomialAnsatz}
n(n\pm2m)a_n+m\tilde{\alpha}a_{n-1}+\varepsilon_{\nu,\pm}^{(n^\prime)} a_{n-2}=0~,
\end{align}
with $a_1=-\tilde{\alpha}m/(1\pm 2m)a_0$, where the upper (lower) sign applies to $m>0$ 
($m<0$). The coefficient $a_0$ is determined by the normalisation condition $2\pi 
\int_0^1d\varrho~\varrho~|\mathcal{F}_{m}^{(\varepsilon_{\nu,\pm}^{(n^\prime)})}(\varrho)|^2=1$.
We note that the polynomial with coefficients given by 
Eq.~(\ref{eq:polynomialAnsatz}) represents a modified Laguerre function that becomes the 
standard Bessel function $J_0(\sqrt{\varepsilon_{\nu,\pm}^{(n^\prime)}}\varrho)$ for $\tilde\alpha=0$ and/or $m=0$. 
The band-edge energies, $\varepsilon^{(n^\prime)}_{\nu,\pm}$ are found by imposing hard
wall boundary conditions on the radial wave function, i.e. for $r=R$, we require
\begin{align}
\mathcal{F}_{m}^{(\varepsilon_{\nu,\pm}^{(n^\prime)})}(\varrho=1)=0~.
\end{align}

\begin{figure}[t]
\includegraphics[width=6.8cm]{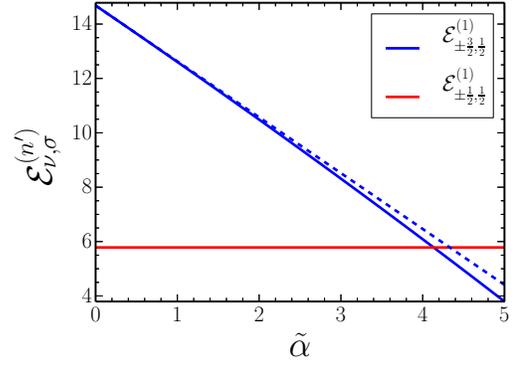}
\caption{\label{fig:Bandedge} 
Energy eigenvalues of the lowest two doubly degenerate quasi-1D subband edges,
plotted as a function of the effective Rashba spin-orbit-coupling parameter $\tilde\alpha$.
The blue dashed curve is an approximation based on Eq.~(\ref{eq:ApproxBandedge}).}
\end{figure}

For not too large values of $\tilde\alpha$, the lowest spin-$\uparrow$ ($\downarrow$)
subband-edge state has $\nu=1/2$ ($-1/2$) total angular momentum. However, as seen from
Fig.~\ref{fig:Bandedge}, a level crossing occurs for $\tilde\alpha\approx 4.2$, beyond which
the new lowest spin-$\uparrow$ ($\downarrow$) subband edge is a state with $\nu=\pm3/2$ ($-3/2$).
The variation of the band-edge energy $\varepsilon_{\pm\frac{3}{2},\pm}^{(1)}$ as a
function of $\tilde\alpha$ can be approximated using standard perturbation theory, yielding
\begin{align}\label{eq:ApproxBandedge}
\varepsilon_{\pm\frac{3}{2},\pm}^{(1)}=\varepsilon_0-\tilde\alpha\,
\frac{\int_0^1~d\varrho~J_1^2(\varrho\sqrt{\varepsilon_0})}{\int_0^1~d\varrho~\varrho~J_1^2
(\varrho\sqrt{\varepsilon_0})}~,
\end{align} 
where $\varepsilon_0\approx 14.68$ is the band edge energy of the corresponding band for $\tilde\alpha=0$.

\section{Regularisation of the integrand for calculating the exchange energy\label{app:MQM}}
In the calculation of the exchange energy we have to deal with integrals of the form
\begin{equation}\label{eq:MQMsetup}
{\mathcal I}=\iint dkdk' G(k,k')K_0\left(|k-k'|\sqrt{r^2+r'^2-2r r'\cos\varphi}\right),
\end{equation}
with $G(k,k')$ being a smooth function of $k$ and $k'$. 
A logarithmic singularity occurs when the argument of $K_0(\cdot)$ vanishes. This happens when either the square root is zero, 
at $\vec{r}_\perp=\pvec{r}'_\perp$, or when $k=k'$. 
To regularise the integral for the case where $\vec{r}_\perp=\pvec{r}'_\perp$, we add a small amount $0^+$ to the term under the square root.
Then by decreasing the value of $0^+$, we perform a series of calculations until the result for the exchange energy doesn't change within a certain tolerance.

The situation for $k=k'$ can be regularised analytically.
To this end, we add to and subtract from Eq.~(\ref{eq:MQMsetup}) the term 
\begin{equation}
\iint dkdk' G(k,k)\ln\left(|k-k'|\sqrt{r^2+r'^2-2r r'\cos\varphi+0^+}\right)~.
\end{equation}
Adding this term to Eq.~(\ref{eq:MQMsetup}) cancels the logarithmic singularity.
The $k'$-integration of the subtracted term can be performed analytically and Eq.~(\ref{eq:MQMsetup}) becomes
\begin{widetext}
\begin{align}\label{eq:MQMIntegrand}
{\mathcal I}=\int dk &\Biggl\{\left[\int dk' G(k,k')K_0\left(|k-k'|\sqrt{r^2+r'^2-2r r'\cos\varphi}\right) +G(k,k)\ln\left(|k-k'|\sqrt{r^2+r'^2-2r r'\cos\varphi}\right)\right] \Biggr.\nonumber\\
&\Biggl.-G(k,k) \left[ k\ln\left(\frac{k_{\text{F}}+k}{k_{\text{F}}-k}\right)-2k_{\text{F}}+2k_{\text{F}}\ln\left(\sqrt{k^2_{\text{F}}-k^2}\sqrt{r^2+ r'^2-2r r'\cos\varphi+0^+}\right)\right]\Biggr\}~.\nonumber\\
\end{align}
\end{widetext}
The expression Eq.~(\ref{eq:MQMIntegrand}) is manifestly finite for $k=k'$.

%%%%%%%%%%%%%%%%%%%%%%%%%%%%%%%%%%%%%%%%%%%%%%%%%%%%%%%%%%%%%%%%%
%\bibliography{ExNanWires}


%

\end{document}